\documentclass{ws-p8-50x6-00}

\begin{document}

\title{Open Beauty Production at HERA}

\author{J. Kroseberg}

\address{{\rm (on behalf of the H1 and ZEUS collaborations)}\\ Physik--Institut der Universit\"at Z\"urich\\ Winterthurer Str. 190,  
CH--8057 Z\"urich, Switzerland\\ 
E-mail: Juergen.Kroseberg@desy.de}

\maketitle

\abstracts{
Recent results on open beauty production 
in positron--proton collisions
at a centre--of--mass energy of 300~GeV
are presented. 
The beauty photoproduction cross section is measured 
by ZEUS and H1 
via the semileptonic decay. 
Both experiments use 
the transverse lepton momentum with respect to
a jet as 
an 
observable. The H1 central silicon vertex detector
makes it possible  
to analyse, in addition, the lepton impact parameter spectrum,
thus exploiting 
lifetime information 
 for the first time at HERA.
The combined lifetime and transverse momentum signature is
further used to observe  
beauty production 
for the first time  
in  
deep inelastic scattering. 
All results 
are compared with NLO QCD calculations.
}

\section{Introduction}

The production of heavy quarks at HERA provides a valuable tool to
probe proton structure and the mechanism of the hard subprocesses
underlying $ep$ interactions. 
Within the framework of  
QCD, heavy quark production is predominantly
gluon induced,  
which allows the gluon content of the proton to be investigated.
The leading--order (LO) process is 
{\it boson gluon
fusion}, where  a photon emitted by the 
positron 
and a gluon coming from 
the proton form a quark-antiquark-pair. 
The exchange of a quasi--real photon  ({\it photoproduction}, $\gamma p$)
dominates over  
large photon virtualities ({\it deep
inelastic scattering}, DIS).
Perturbative QCD (pQCD) calculations, assuming the $b$ quark to be dynamically
produced in the hard subprocess ({\it massive} approach), are available
in next--to--leading order (NLO). In the case of beauty production, pQCD predictions are expected
to be particularly reliable because the $b$ quark mass sets a hard
scale. For this reason, $b$ production is considered an excellent testing
ground for pQCD 
even though the cross section is two and three orders of
magnitude below 
charm production and total cross sections, respectively.
The first measurement
of open $b$ photoproduction at HERA\cite{h1openb} yielded a  cross
section significantly above the QCD expectation. A similar discrepancy between
data and theoretical prediction   
has been observed in
$p\bar{p}$\cite{hadronb} 
and, recently, in  $\gamma\gamma$\cite{ggb} 
interactions. 
New data and improved experimental tools have now become available to further
study 
$b$ production processes at HERA.
Recent analyses,   
based on semileptonic $b$ hadron 
decays within jets, are presented 
here.

\section{ZEUS}
\begin{figure}[h]
\vspace*{-0.5cm}
\epsfxsize=11.2pc %
\epsfbox{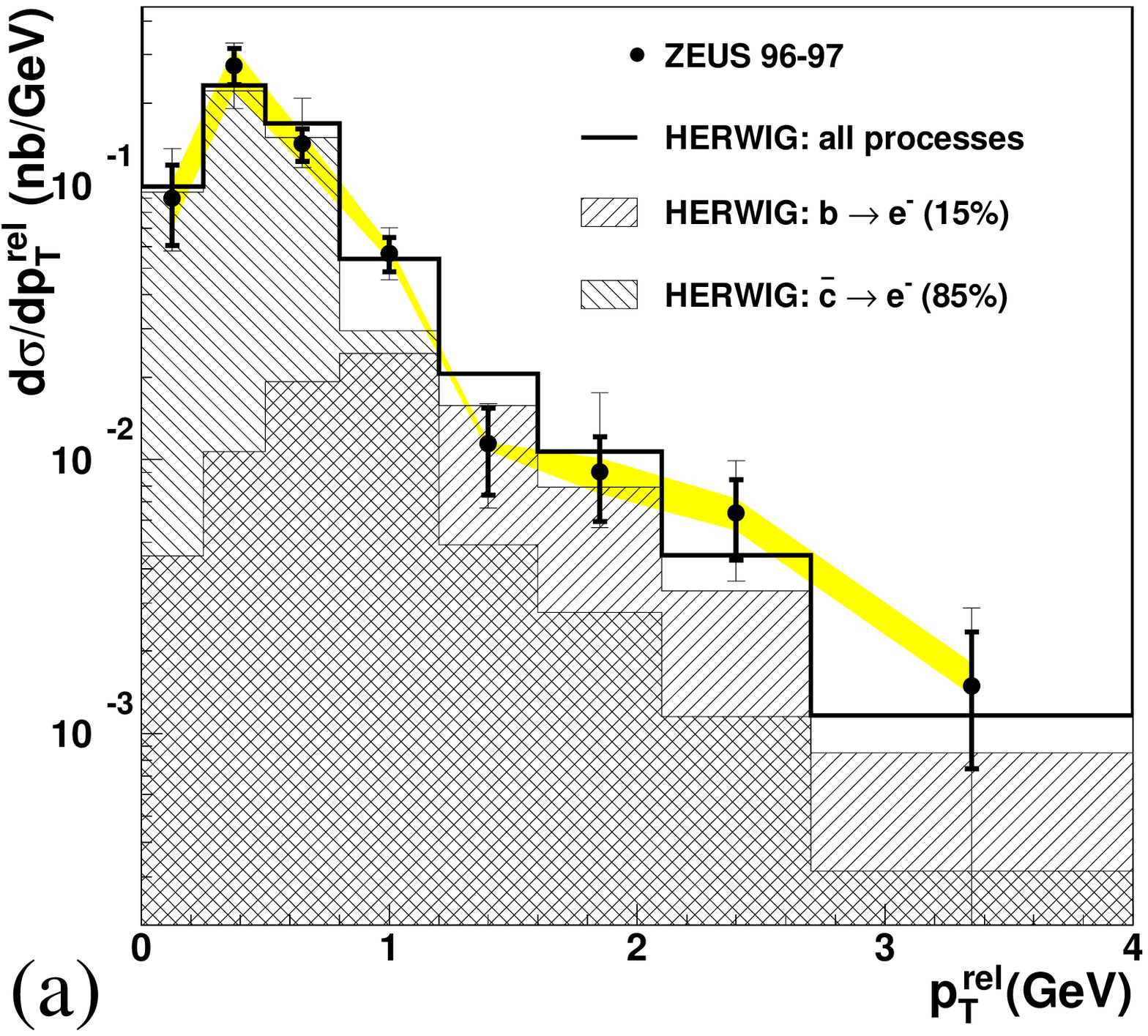} %
\epsfxsize=15pc
\hspace*{0.3cm}\epsfbox{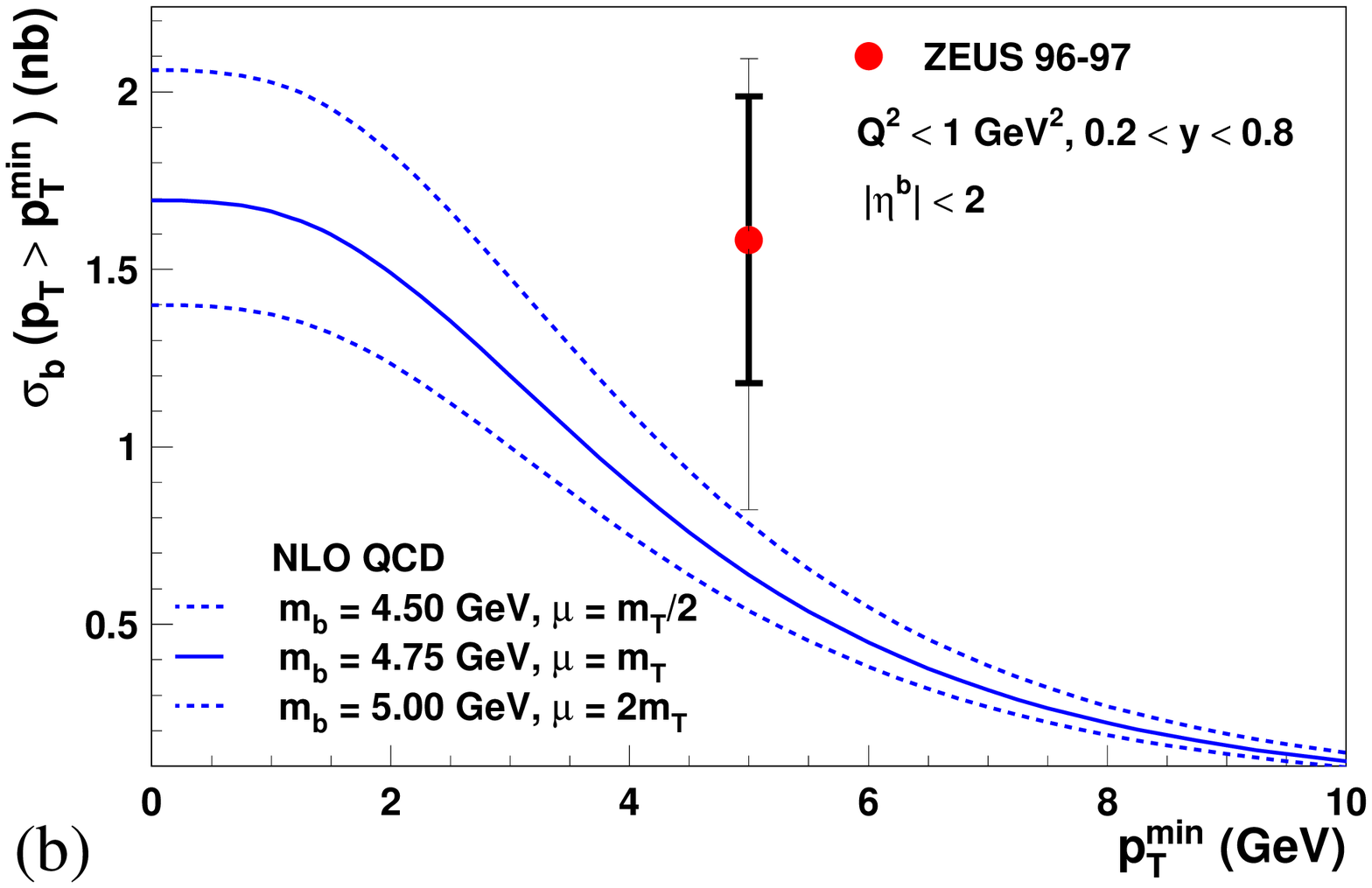} %
\vspace*{-0.4cm}
\caption{(a) $d\sigma/dp_t^{rel}$ for data and MC (normalised and fitted
to the data).\newline
(b) Extrapolated $\sigma$ from data and NLO QCD (as a function of the minimum $p_t^b$). 
\label{fig:ZEUSptr}}
\end{figure}

ZEUS 
has measured $b$ photoproduction
in inclusive muon\cite{zeusmuon} and electron analyses. 
Here only the electron analysis is discussed, which has recently been
published\cite{zeusopenb}. 
From data 
collected during
1996 and 1997, which correspond  
to an integrated luminosity of \mbox{${\cal
L}=38.5$~pb$^{-1}$},   
 a $\gamma p$ sample is obtained by
rejecting events  
in which 
the scattered positron is found in the main
detector. 
Jets are reconstructed, and events with 
at least two jets with transverse energy 
$E_t(\mbox{jet}_{1(2)})>7(6)$ GeV are selected. 
Electrons are
identified 
by 
combining calorimeter and drift chamber information. The
hadronic background is statistically subtracted using the specific
energy loss  
in the drift chamber.  
Backgrounds from Dalitz
decays and photon conversions are removed explicitly, resulting in a sample
of ($943\pm 69$) electrons with a transverse momentum $p_{t}>1.6$ GeV 
and pseudorapidity $|\eta| < 1.1$.
For each electron, the transverse momentum
 $p_t^{rel}$ relative to the closest jet is reconstructed.
Due to the high $b$ mass 
a harder $p_t^{rel}$ spectrum is expected for
 the signal compared to the background. 
The measured differential cross section, $d\sigma/dp_t^{rel}$, is shown in
 figure~\ref{fig:ZEUSptr}(a), together with the  prediction for beauty
 and charm production from the LO Monte Carlo program \mbox{HERWIG\cite{herwig}},  
 including contributions from resolved $\gamma$ processes.
Taking the normalisation from the data, a maximum likelihood fit  
results in a beauty fraction $f_b=[14.7\pm 3.8(stat.)]\;\%$, which
translates  into a visible cross section of $[24.9 \pm 6.4(stat.) ^{+4.2}_{-7.3}(syst.)] \ {\rm pb}$. An 
 extrapolation to the parton level for the region $p_T^b>$ 5 GeV,
 $|\eta^{b}|<2$, photon virtuality 
 $Q^2<1\;{\rm GeV^2}$ and inelasticity  $0.2<y<0.8$ gives
$$\sigma^{\rm ext}_{\gamma p} = [ 1.6 \pm 0.4 (stat.) ^{+0.3}_{-0.5} (syst.) ^{+0.2}_{-0.4}
(extrapolation) ]~\mbox{nb},$$ 
somewhat above  the result of a  NLO QCD calculation,\cite{frixi} 
cf. figure \ref{fig:ZEUSptr}(b). 

\section{H1}

\begin{figure}[h]
\vspace*{-0.8cm}
\epsfxsize=13pc %
\epsfbox{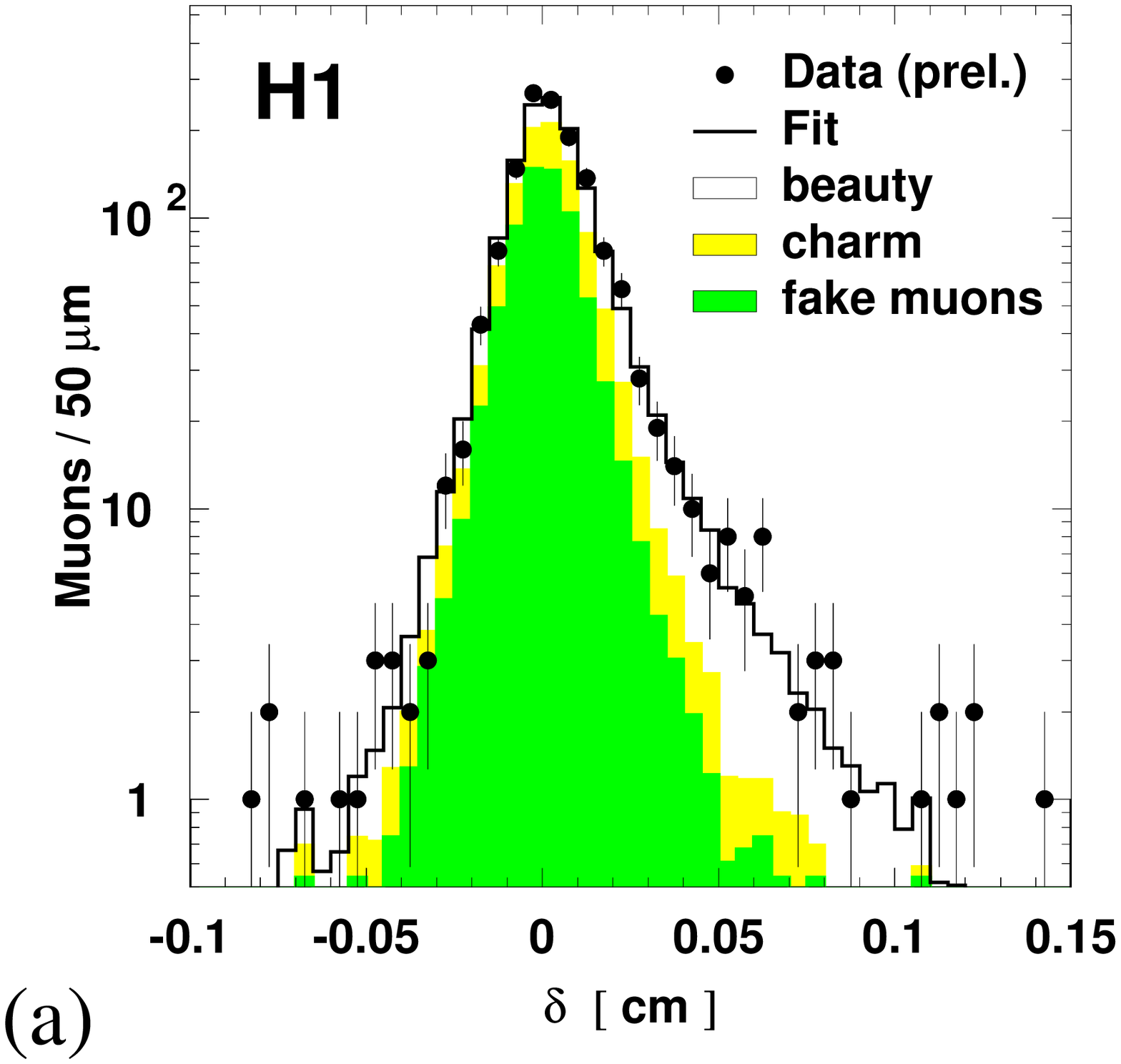} %
\epsfxsize=13pc
\hspace*{0.8cm}\epsfbox{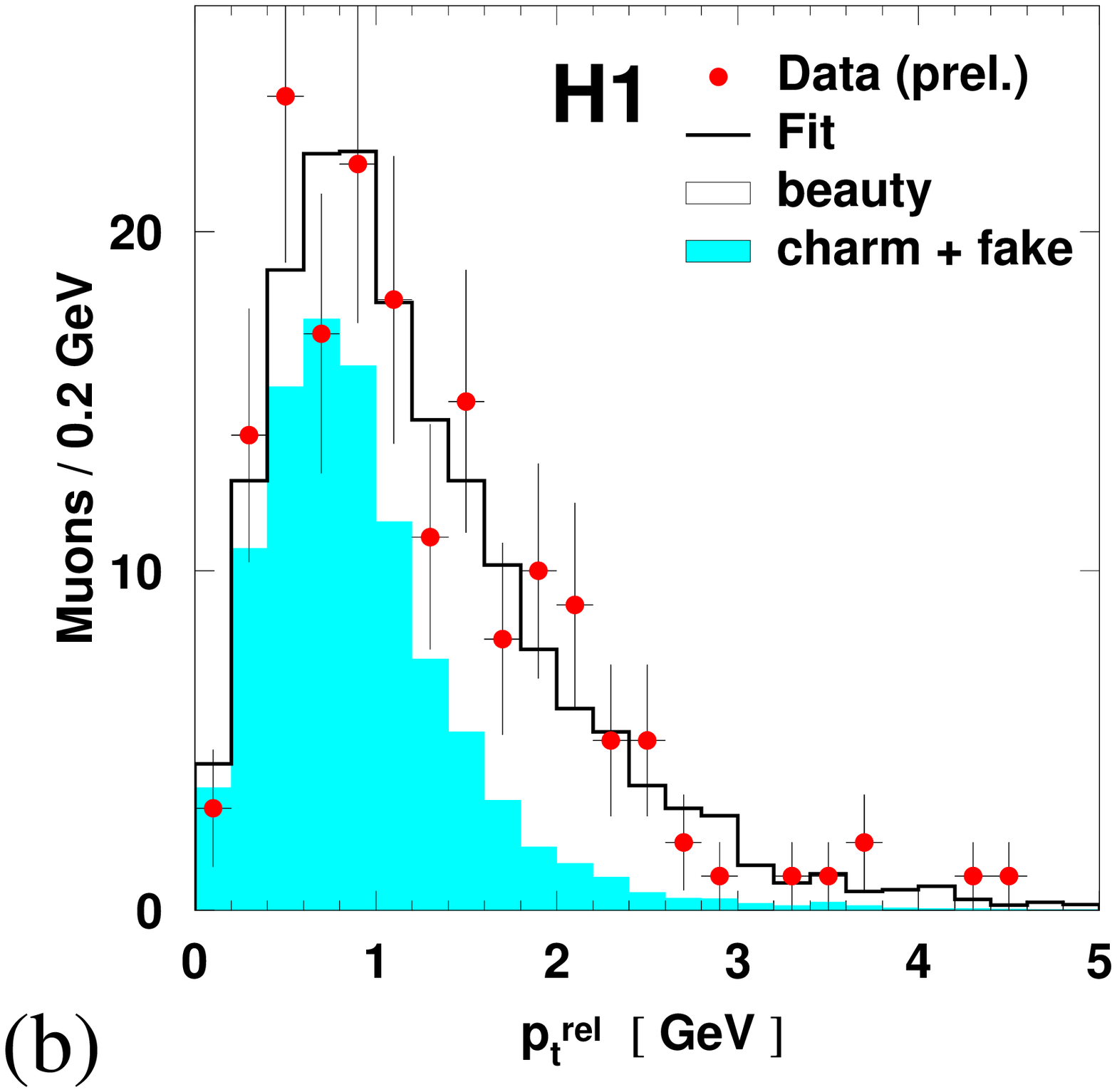} %
\vspace*{-0.7cm}
\caption{(a) Muon $\delta$ distribution ($\gamma p$ sample) with
the  decomposition from
the $\delta$ fit. \newline  
(b) $p_T^{rel}$ spectrum for the DIS sample, 
with the result of the combined ($\delta$,
$p_T^{rel}$) fit.
\label{fig:H1}}
\vspace*{-0.3cm}
\end{figure}

The first H1 measurement of open $b$ production\cite{h1openb} was based on 
an inclusive muon $p_t^{rel}$ analysis using 1996 $\gamma p$ data. 
In the new analysis, the H1 central silicon tracker CST\cite{cst}
is used to detect $b$ hadrons also via their long lifetimes. 
In a sample of dijet $\gamma p$  events 
($E_t(\mbox{jet}_{1,2})>5$ GeV) 
collected in 1997 \mbox{(${\cal
L}=14.7$~pb$^{-1}$)},  1415 muon candidates are selected, which 
are  identified in the 
instrumented iron and  the central drift chamber  
and precisely measured
in the CST.
The decomposition of the sample is obtained
from a likelihood fit to 
the muon signed impact parameter ($\delta$) spectrum,\footnote{
$\delta$ is reconstructed in the plane perpendicular to the beam.
$|\delta|$ is 
the distance
of closest approach of the candidate track to  the event primary
vertex. The sign is positive if  the track crosses the 
jet axis downstream with respect to the vertex, and negative
otherwise. 
}
which at large positive values is expected to be signal dominated.
Contributions from beauty and charm 
production are modelled by the AROMA
MC program\cite{aroma} based on LO QCD and parton showers. 
Light quark background, dominated by
misidentified hadrons ({\it fake muons}), is selected from data.
The fit describes the data well, cf. figure \ref{fig:H1}(a), and yields 
a signal contribution of 
$f_b = (26\pm 5(stat.))\,\%$. Translating this into a visible cross section,
the previous H1 measurement is confirmed using  
new data and an independent signature. 

After having verified 
the consistency of 
$p_t^{rel}$ and  $\delta$, the separation power of both
observables is combined in a fit to the two--dimensional \mbox{($\delta$,
$p_T^{rel}$)} distribution. The resulting cross section in the visible
range, defined by \mbox{$p_t(\mu)>2$ GeV},
$35^\circ<\theta(\mu)<130^\circ$, 
$Q^2<1$ GeV$^2$ and $0.1<y<0.8\,$, is
\[ \sigma_{\gamma p}^{vis}=
[160\;\pm\;16\;(stat.)\; \pm 29\;(syst.)\;]\;{\rm pb}\ . \]
The combination of published and
new results yields a visible cross section of $[170
\pm 25] \; {\rm pb}\;$.
A NLO QCD calculation with the FMNR program\cite{frixi} gives a 
significantly lower result of 
$[54\;\pm\;9\;]\;{\rm pb}$.\cite{h1openb}

Having established the new method in the photoproduction regime, it 
is further used to measure for the first time the open $b$ production
cross section in DIS.
By reconstructing the scattered beam positron in the backward calorimeter,
a  
DIS event sample is obtained, which corresponds to an integrated
luminosity of ${\cal
L}=10.5$~pb$^{-1}$. After applying the same muon and jet
selection as in the photoproduction analysis and restricting the
kinematic range to $2$ GeV$^2<Q^2<100$ GeV$^2$ and $0.05<y<0.7\,$, 171
muon candidates remain. The combined ($\delta$,$p_T^{rel}$) fit 
results in $f_b =
(43\pm 8(stat.))\,\%$ and gives a good description of the data; for the 
$p_T^{rel}$ projection this is shown in figure
\ref{fig:H1}(b).\linebreak  
The
corresponding visible cross section is 
\[\sigma_{\rm  DIS}^{vis} = [39\;\pm\;8\;(stat.)\;
\pm 10\;(syst.)\;]\;{\rm  pb}.\]
This again exceeds significantly the NLO QCD prediction 
of 
$(11 \pm 2)\;{\rm pb}$ 
obtained from the HVQDIS program.\cite{hvqdis}

\vspace*{-0.15cm}
\section{Summary}
Open beauty production is measured in inclusive lepton
analyses by ZEUS and H1 using new data and 
including
the  $b$ lifetime signature. 
Previous photoproduction results are 
confirmed and  
improved, 
and the DIS cross section is measured for the first time.
All measured cross sections are found 
to be above 
the NLO QCD predictions. For the H1 results the excess is significant.

\end{document}